# Comparative study of P and S wave amplitudes of acoustic logging through solid formations : contribution to the knowledge of *in situ* stresses and fractures

André Rousseau

*CNRS-UMS 2567 (OASU)*
*Université Bordeaux 1 - Groupe d'Etude des Ondes en Géosciences*
*351, cours de la Libération, F-33405 Talence cedex*
a.rousseau@geog.u-bordeaux1.fr

**Abstract**

In order to compare the ratios between the amplitudes of P and S acoustic waves drawn from various boreholes, we have defined a specific wave parameter called magnitude : for each shot, we have added the maximum positive amplitude values and the absolute values of the minimum negative amplitudes of a given wave recorded at different depths.

In compact formations, the difference between the logarithms of the magnitudes of S and P waves provides a value that appears to be characteristic of the state of stress of the domain where a borehole is drilled. Accidents between those formations, such as fractures or sedimentary joints can be determined, and breakouts well distinguished.

In order to compare the ratios between the amplitudes of P and S acoustic waves drawn from various boreholes, we have defined a specific wave parameter called magnitude : for each shot, we have added the maximum positive amplitude values and the absolute values of the minimum negative amplitude of a given wave recorded at different depths.

In compact formations, the difference between the logarithms of the magnitudes of S and P waves provides a value that appears to be characteristic of the state of stress of the domain where a borehole is drilled. Accidents between those formations, such as fractures or sedimentary joints can be determined, and breakouts well distinguished.

**Introduction**

The variation of the amplitudes of the acoustic waveforms propagating through a borehole has been taken into consideration for several decades in order to locate open fractures and estimate porosity. Stoneley waves supply good information for the first case and body waves for the second one. However, the quantitative comparison between the simultaneous amplitude variations of P and S waves is not a usual parameter.

In fact, it is mainly the apparent difficulty to obtain P waves of sufficient amplitudes in acoustic logs of some boreholes through basalts which questions the mechanisms of propagation. A series of articles was edited by Williamson (2003) concerning the seismic imaging problems of the sub-basalt sequences, but the wave frequencies in both techniques (reflection seismic and acoustic logging) are not of the same order (10 Hz / 20 kHz). This paper proposes to calculate the ratios between the amplitudes of P and S waves around several deep boreholes and to compare the results between them in relation to the local geological environment. This work makes sense only in the case of solid formations.

**Protocol followed to calculate the "magnitudes" of the P and S waves**

Eight or two receivers of sonic monopole probes provide the data, that is to say there are eight or two waveforms recorded for each shot. For each wave at each depth, we have added the maximum positive amplitude value and the absolute value of the minimum negative amplitude. *The sum of the eight or two such values for a given wave characterizes what we call the wave magnitude at the given depth* (Figure 1).

As the aim of this calculation is the comparison between the results from different boreholes, the question of the dynamics of the signals arises, and it cannot be directly resolved. When a resonance occurs (see Rousseau, 2005a), the calculated magnitude may reflect a parameter different from the case without resonance. Finally, one must remember that open fractures decrease the amplitudes of body waves, as well as those of surface waves (Rousseau and Jeantet, 1997).

On the other hand, as we take into consideration only solid formations in this work, and thanks to large transmitter-receiver distances (more than 3 meters), one can easily separate automatically P and S waves from their respective velocities. However, it is very difficult to separate S waves from their corresponding surface waves, the pseudo-Rayleigh waves that are dispersive and arrive between the S and direct waves. However, their maximum amplitudes are often similar.

The use of the logarithm of the so-called magnitudes allows us (i) to be free of the signal dynamics in order to compare the P and S waves of the waveforms at each depth, and (ii) to smooth the variations of the results, particularly in the case of S waves. So *the difference between the logarithm of S wave magnitude* (called Sw) *and the logarithm of P wave magnitude* (called Pw), which is a ratio, will be the parameter chosen for the comparison between different wells. This parameter is characteristic for a given frequency because it does not depend on signal processing nor data acquisition conditions.

**Case histories**

We have calculated the ratios between P and S waves of monopole acoustic logs (about 20 kHz) from more or less deep wells through various solid rocks. They are :
- the KTB Pilot Hole in Bavaria (Germany) drilled through gneisses and amphibolites,
- the GPK1 borehole (Soultz sous forêt in Alsace, France) drilled through granites,
- the SAFOD Pilot Hole (California, USA) drilled through granites,
- the Auriat borehole (Massif Central, France) drilled through granites,
- the ANDRA boreholes drilled in the Vienne region (France) through granites,
- the Balazuc1 borehole (south of France) drilled through sandstones, limestones, dolomites and calcarenites,
- the ODP Hole 735b (LEG 176) drilled through gabbros of the SW Indian Ridge (SW Indian Ocean),
- the ODP Hole 1137A (LEG 183) drilled in the Kerguelen Plateau through basalts (Antarctic).

**Analysis of the results**

Plate I displays the logarithms of the P wave and of the S wave magnitudes, called respectively Pw and Sw, with their differences called Sw – Pw, for all the mentioned boreholes with a diagram for each run. The logarithms of the Stoneley wave magnitudes are plotted for information. The results can be examined under two main approaches : 1) compact formations, and 2) fractures or joints between compacts formations.

   a) **within compact formations**

The mean values of Sw - Pw vary only slightly between "accidents" which generally cause decreases (see next section). These values are indicated below.

- For the **KTB Pilot Hole** : from **2.0** at *4000 meters* of depth up to **3.0** at *890 meters* ; as the well crossed a sub vertical fracture near the surface, the low values from the runs KTB_P 1 and KTB_P 4 are not reliable for characterizing compact formations.
- For the **GPK1 borehole (Soultz sous forêt)** : from **2.5** at *3500 meters* of depth to **3.8** at *2200 meters*.
- For the **SAFOD Pilot Hole** : from **1.5** at *800 meters* of depth to **2.0** at *2100 meters*.
- For the **Auriat borehole** : from **2.5** to **3.0** between *900* and *540 meters* of depth. The respective offsets are 3.20 and 5.68 meters for the small and great size tools.
- For the **ANDRA boreholes** :
    - **CHA106** : from **0.5** at *180 meters* of depth to **1.0** at *550 meters*,
    - **CHA105** : between **1.5** and **2.0** from *210* to *350 meters* of depth,
    - **CHA112** : **1.75** at *180 meters* of depth to **2.0** at *570 meters*.
- For the **Balazuc1 borehole** : **1.4-1.5** between *1700 meters* of depth and *600 meters*, with a passage up to **2.75** between *1000* and *650 meters*.
- For the **ODP Hole 735b** : **3.4** between *100* and *580 meters* of depth.
- For the **ODP Hole 1137A** : **2.0** between *230* and *350 meters* of depth. The weak values, here not representative, are consecutive to tool vibrations or sedimentary "joints" between basalts.

Figure 2 recapitulates the values of Sw-Pw. We observe that the wells drilled within known stress domains are spread out so that those in a tensile domain – the Soultz borehole (Cornet et al, 1997; Genter et al, 1997; Klee and Rummel, 1999) and the KTB Pilot Hole (Bücher et al, 1990; Zang et al, 1990; Roeckel and Natau, 1993; Rousseau, 2005a) – correspond to larger values than this in a compressive and shear domain – the SAFOD borehole (Boness and Zoback, 2004; Chéry et al, 2004; Korneev et al, 2003; Townend and Zoback, 2004; Rousseau, 2005b).

The position of the Auriat borehole indicates a low stress domain within the French Massif Central, while the lowest values of Sw-Pw of the Balazuc1 borehole tend to belong to a stressed domain in the Cevennes Massif. The dashed line in Figure 2 indicates higher values probably due to a resonance of the S waves propagating within calcarenites. Thus, those values should represent this kind of sedimentary formation rather than the local stress. As for the ANDRA boreholes in the much fractured Vienne region in France (Gros and Genter, 1999), the low values of Sw-Pw clearly suggest a stressed domain, which had already been deduced by the author (Rousseau, 2005a). Resonance of P waves might cause the very weak Sw-Pw values of CHA106, unless the juxtaposition of two stressed domains of unequal magnitude might produce the differences of the Sw-Pw values between CHA106 on the one hand, and CHA105 and CHA112 on the other hand, although the distance between them is small (less than 2 km). In that case, this area might be the location of a future rupture.

In this frame, the locations of our ODP wells relative to the Sw-Pw values are interesting : the Hole 737b drilled over the Indian rift is characteristic of a tensile domain, while the Hole 1137A drilled within the Kerguelen Plateau is in an intermediate position, between tension and compression.

When the vertical logged segment of a well is long enough, we can observe some variations of the Sw-Pw values. We have noted the influence of some sedimentary formations, such as calcarenites in the Balazuc1 borehole, but we observe monotonic variations with depth in crystalline formations. The Sw-Pw values decrease with depth in the KTB Pilot Hole and the Soultz borehole, while they increase in the SAFOD Pilot Hole. In the first case, the stress would increase with depth, and in the second case, the stress would decrease.

**b) case of fractures or joints between compact formations**

Fractures and joints reveal a decreasing of Sw, sometimes simultaneously with Pw. In the diagrams of the Plate I, we have plotted the logarithm of the corresponding Stoneley wave magnitudes Stw for information.

In the case of an ***open fracture***, we observe that Sw, Pw and Stw decrease simultaneously. It is the only case of Stw decreasing. If Sw decreases, but Stw does not, while Pw may sometimes increases simultaneously because of resonance, the occurrence of ***breakouts*** may cause this phenomenon. It is particularly the case in the KTB Pilot Hole (see Kück, 1993). ***Sedimentary joints***, as those in the ODP Hole 1137A, cause a resonance, inferring the increasing of Pw. Finally, Pw and Sw may increase because either ***friction of the tool*** against the hole wall (case in the Balazuc1 borehole) or ***resonance of the tool*** itself (it is perhaps the case in the ODP Hole 1137A). It is unusual to observe only Pw decreasing.

**Discussion and conclusion**

In the case of compact and no porous formations, the study of the ratio between the maximum values of the amplitudes of each acoustic body waves shows that this ratio is not a tool parameter, but the consequence of an environmental parameter. The examples of various geological regions allow us to determine it as the *in situ* stress. In this case, the values Sw-Pw ought to be used as a scale representative of the kind of stress domain : the lower they are, the higher the stress is. The shear and compressed domains correspond to values around 1, and the very weak stress domains to values up to 3.5.

The simultaneous variations of the values Sw, Pw and Stw provide invaluable information about the "accidents" affecting those formations, such as fractures or all kinds of joints. In addition to usual well images, we have an easy means to distinguish breakouts from fractures. The vertical axial position of those induced cracks causes the resonance of P waves, but the attenuation of S waves.

The physical interpretation of those results is that S waves vibrate as less as stress is great. The value Sw-Pw represents therefore a kind of *freedom degree of the transversal vibration* of S waves, characteristic of the crossed formation.

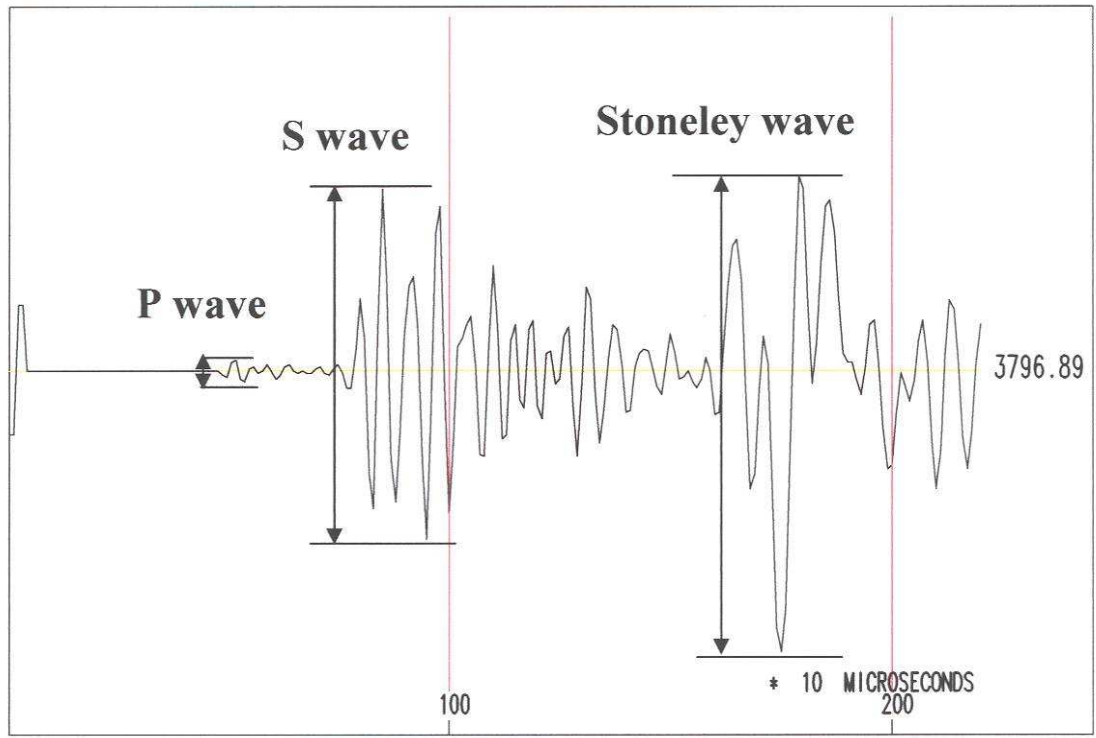

Figure 1 : Wave magnitudes.
The magnitude of each kind of wave is calculated from the amplitudes indicated by arrows.

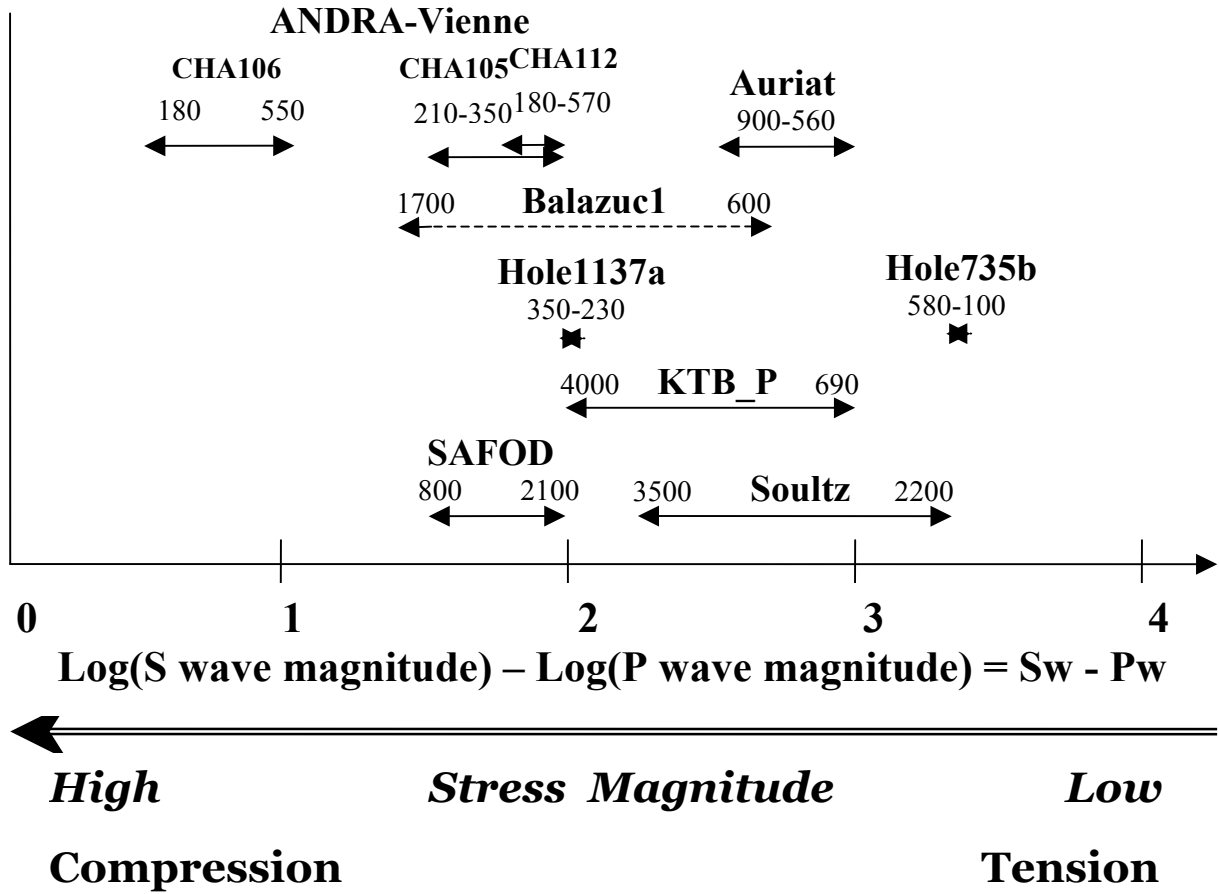

Figure 2 : Diagram recapitulating the values Sw-Pw of all the studied boreholes.
The two numbers next to each borehole name indicate depths in meters.